\documentclass[10pt, conference, letterpaper]{IEEEtran}

\usepackage{tikz}
\usepackage{amsmath}
\usepackage[T1]{fontenc}
\usepackage[utf8]{inputenc}
\usepackage{graphicx}
\usepackage{epsfig}
\usepackage{amsmath}
\usepackage{bbold}
\usepackage{color}
\usepackage{algorithm}
\usepackage{algorithmic}
\usepackage{wrapfig}
\usepackage{booktabs}
\usepackage{pifont}
\usepackage{xspace}
\usepackage{cite}
\usepackage{url}
\usepackage{graphicx}
\usepackage{subcaption}
\usepackage{booktabs}
\usepackage{adjustbox}
\usepackage[normalem]{ulem}
\usepackage{amssymb}

\graphicspath{{./figs/}}

\newtheorem{constraint}{\textbf{Constraint}}

\newcommand{\sysname}{zkSAS\xspace}

\title{zkSAS: Practical Zero-Knowledge Proofs for Verifiable Spectrum Access Management}

\author{
\IEEEauthorblockN{
Nishat F. Purbasha\IEEEauthorrefmark{1},
Ifteher Alom\IEEEauthorrefmark{1},
Eric W. Burger\IEEEauthorrefmark{2},
Y. Thomas Hou\IEEEauthorrefmark{2},
Wenjing Lou\IEEEauthorrefmark{2},
Yang Xiao\IEEEauthorrefmark{1}
}

\IEEEauthorblockA{\IEEEauthorrefmark{1}University of Kentucky, Lexington, KY, USA}
\IEEEauthorblockA{\IEEEauthorrefmark{2}Virginia Tech, Blacksburg, VA, USA}
}

\begin{document}
\maketitle

\begingroup
\renewcommand\thefootnote{}
\footnotetext{
This is the authors' final version of the paper, which has been accepted for publication at the 2026 IEEE International Symposium on Spectrum Innovation (DySPAN), Washington, DC, USA, 2026.
}
\addtocounter{footnote}{-1}
\endgroup

\begin{abstract}

Dynamic Spectrum Access (DSA) through the Spectrum Access Systems (SAS) elevates spectral efficiency, yet existing centralized models face allocation logic opaqueness and a lack of independent verifiability. While blockchain-based SAS architectures offer transparency and verifiability by default, they introduce critical privacy risks and prohibitive on-chain computational overhead. We introduce zkSAS, a practical zero-knowledge proof (ZKP) system designed to address the verifiability and privacy gaps in SAS deployments, with direct applicability to both the existing CBRS SAS model and blockchain-based SAS models. The system features a suite of ZKP circuits, encompassing proofs of allocation constraint validity and proofs of move list validity to verify that channel assignments and move list-based incumbent protection measures, respectively, adhere to regulatory constraints without exposing sensitive user data. Comprehensive evaluation of our prototype in both centralized and blockchain-based settings indicates that while proof generation scales with spectrum user population, verification remains lightweight and constant-time. We envision that zkSAS offers a scalable and practical path to secure, verifiable dynamic spectrum sharing.


\end{abstract}



\section{Introduction}
\label{sec:intro}


Dynamic spectrum access (DSA) has emerged as a transformative paradigm for optimizing spectral efficiency by opening frequency bands that are traditionally reserved for exclusive incumbent use to opportunistic access by commercial secondary users. As a prominent realization of this concept, the Spectrum Access System (SAS) \cite{fcc2015rule} designed for serving the 3.55-3.7 GHz Citizens Broadband Radio Service (CBRS) band, has seen widespread adoption since 2020. Architecturally, an SAS functions as a centralized coordinator and database, dynamically authorizing spectrum grants for commercial users while protecting the preemptive access rights of incumbents such as naval radars and satellite ground stations. 

However, current SAS implementations face an outstanding verifiability challenge, which stems from the centralized service model characterized by opaque allocation logic. Since the spectrum access right is an intangible asset, its value depends entirely on the validity and enforceability of the grants created by the SAS. Without a mechanism for independent verifiability, secondary users must place blind trust in the SAS for executing complex interference calculations and policy compliance. This lack of transparency and verifiability creates significant legal and commercial risks, as erroneous allocations can lead to destructive interference to incumbent and priority access users or the unfair exclusion of secondary access users. Such faults are difficult to audit in a closed-server architecture. Therefore, establishing the verifiability of SAS operations is essential for the reliability of a spectrum economy where access rights can be trusted and legally defended.


Meanwhile, recent research has explored decentralized, blockchain-based SAS architectures that realize verifiable spectrum management \cite{weiss2019application, ariyarathna2019dynamic,zhang2020blockchain, grissa2019trustsas, xiao2023bd, shi2024trisas}. A blockchain system offers a unique advantage primarily due to its built-in verifiability, transparency, irreversibility, and limited trust assumptions on central service providers \cite{weiss2019application}. While varying in implementations, existing blockchain-based SAS solutions typically employ ``on-chain'' programs, commonly known as smart contracts, to execute channel assignment logic. Some proposals also incorporate on-chain compensation logic and trading mechanisms for short-term band leasing \cite{ariyarathna2019dynamic,xiao2023bd}, echoing the FCC vision that the blockchain technology has the potential to enable a dynamic spectrum economy with light administrative overhead \cite{rosenworcel2018fcc}.

Despite filling the verifiability gap, blockchain-enabled SAS models inevitably bring new challenges in privacy and computational efficiency. 
First, the inherent transparency of blockchain ledgers exposes sensitive operational data, including user device identifier, geo-location, temporal access pattern, etc., to all blockchain observers. The exposed user records also increase the risk of movement tracking on incumbent users. Meanwhile, spectrum allocation algorithms are often considered as intellectual property of SAS service providers; encoding them on-chain erodes business interests.
Second, on-chain spectrum assignment imposes a heavy computational overhead due to the replicated execution model defined by blockchain consensus. Such overhead significantly restricts the complexity of the spectrum allocation logic. Existing algorithms such as the standardized Iterative Allocation Process (IAP) \cite{WIForum-CPAS} and resource optimization-based allocation methods \cite{manosha2017channel,basnet2018fairness,ying2018sas,basnet2019resource,jai2021optimal} would incur prohibitive on-chain costs.

\textbf{Contributions.}
We introduce \textbf{\sysname}, a practical zero-knowledge proof (ZKP) system that fills the critical verifiability gap in SAS deployments. \sysname can be used independently to elevate the trustworthiness of existing SAS models by proving spectrum allocation validity to clients and facilitating regulatory audits. \sysname can also be integrated into blockchain-based SAS to bridge the privacy-efficiency gap by generating off-chain execution proofs. These proofs certify that spectrum allocations produced in an off-chain private environment are valid and serve as direct triggers for any on-chain logic contingent on a verified allocation. 

Concretely, the \sysname system at its core consists of a suite of ZKP constructions designed to prove the validity of two classes of spectrum operational outcomes.

\begin{itemize}
    \item \textbf{Proof of allocation constraint validity.} We construct ZKP circuits that verify whether the channel assignments in a set of grants satisfy a comprehensive list of constraints established in prior work \cite{fcc2015rule,jai2021optimal}. To ensure both privacy and authenticity of allocation parameters binding to the proof, we employ a commit-and-prove design in which the specific allocation parameters are treated as a private witness while their cryptographic commitment is exposed as a public signal. This design ensures that users receive the authentic allocation associated with the proof and enables independent audits by a third party, without exposing sensitive user-specific assignment data. In our implementation, while proving time grows with the total number of counties and users per county encoded in the allocation, verification remains lightweight and nearly constant-time. (Section \ref{sec:sys-design-constraints})
    
    \item \textbf{Proof of move list validity.} We construct ZKPs for Move List \cite{WIForum-IncumbentProtection}, an essential function that protects incumbents by determining which commercial users should stop transmitting. Our proof construction encompasses the validity of key algorithmic checks rather than encoding the entire move list procedure in-circuit. Complex steps such as sorting and interference aggregation are executed off-chain, and their outputs are supplied as private witnesses. Similar to the proof of allocation constraint validity, while proving time grows with the size of the set of grants encoded in the witness, verification remains constant-time. (Section \ref{sec:sys-design-movelist})
\end{itemize}


We provide three use cases of \sysname in the contexts of the existing CBRS SAS paradigm and blockchain-based SAS models, followed by a new use case involving Spectrum Bux for future spectrum sharing (Section \ref{sec:usecase}). We implemented a \sysname prototype and evaluated its computational overhead in both centralized SAS and blockchain-based SAS settings  (Section \ref{sec:eval}). Across all settings, the dominant overhead comes from proof generation, whereas the verification is done efficiently and largely insensitive to the spectrum user population. Overall, the measured end-to-end latency stays practical at realistic county- and user-scale workloads, and it scales in a predictable way as the user population increases.

\section{Preliminaries}
\label{sec:background}


\subsection{Spectrum Access System Basics}

\textbf{SAS service model.} Existing SAS implementations for the CBRS band follow a centralized client-server architecture. A spectrum user must register its devices, formally known as CBRS Devices (CBSDs), with a certified SAS administrator, accompanied by critical operational parameters such as geographic coordinates, antenna height, and device category \cite{fcc2015rule}. Following registration, the user performs an \texttt{Inquiry} to identify available channels in its vicinity. To utilize a specific channel, the user then submits a \texttt{Grant} request; the SAS evaluates this request against the current interference environment and incumbent protection requirements, which is typically executed by a spectrum allocation algorithm (see below). Once a grant is issued, the device remains silent until it begins to periodically send out \texttt{Heartbeat} messages to the SAS. This periodic signaling mechanism allows the SAS to dynamically manage the spectral environment and authorize the CBSD for radio transmission at a short time scale. 


\textbf{Allocation algorithms.} The generation of grants typically involves an allocation algorithm. The standard baseline is the Iterative Allocation Process (IAP), which iteratively evaluates the feasibility of each grant request by calculating aggregate interference and ensuring it remains below defined protection thresholds. While SAS administrators are permitted to use their proprietary allocation algorithms, the allocation result should be substantively similar to IAP. In current commercial deployment, allocation is performed during the nightly synchronization windows mandated by the Coordinated Periodic Activities among SASs (CPAS) procedure \cite{WIForum-CPAS}. 

\textbf{Incumbent protection with move list.} A vital mission of SAS is to protect the pre-emptive access rights of incumbents. An incumbent, such as a naval radar, can arrive at any location in one of the Dynamic Protection Areas (DPAs). A SAS detects incumbent arrivals at DPAs via its Environmental Sensing Capability (ESC) units, which are deployed at fixed DPA points that reside at the border of DPAs. The SAS is required by law to suspend commercial grants, and signal affected users to stop transmitting in incumbent-occupied channels within five minutes \cite{WIForum-IncumbentProtection}.
To facilitate the determination of grant suspension, the \emph{move list} algorithm is standardized \cite{WIForum-CommercialOp} following the principle that higher interference creators are first to be put into the move list and have their grants suspended; the remaining grants are not affected.


\subsection{List of Spectrum Allocation Constraints}
\label{subsec:list-constraints}

A spectrum allocation specifies the channel assignments to users by an SAS.
We summarize the spectrum assignment notations and allocation constraints according to Jai et al. \cite{jai2021optimal}. Formally, an \emph{allocation} is defined as a set of channel assignments to all registered users (extracted from the grants): 
\begin{equation*}
    \mathbf{alloc}:= \big\{\big\{x_{i|j}^c\big\}_{i\in\mathcal{P}_j,c\in\mathcal{F}_\mathrm{P}},\big\{y_{i|j}^c\big\}_{i\in\mathcal{G}_j,c\in\mathcal{F}}\big\}_{j\in\mathcal{A}}
\end{equation*}
where $x^c_{i|j}\in\{0,1\}$ is the binary indicator for PAL channel assignment---it equals $1$ if channel $c$ is assigned to PAL user $i$ for county $j$. Similarly, $y^c_{i|j}\in\{0,1\}$ indicates whether channel $c$ is assigned to GAA user $i$ for county $j$. $\mathcal{P}_j$ is the set of PAL users in county $j$; $\mathcal{G}_j$ is the set of GAA users in county $j$. $\mathcal{F}_\mathrm{P}$ is the set of PAL-eligible channels (i.e., channels 1 to 10 in the CBRS band); $\mathcal{F}$ is the set of all channels. Lastly, $\mathcal{A}$ is the set of counties considered in this allocation.


\begin{constraint}[PAL Assignment per Channel-and-County]
At most one PAL user can be assigned to each PAL-eligible channel in any county. That is,
\begin{equation}
\label{eq:constraint1}
    \forall c\in\mathcal{F}_\mathrm{P},\forall j\in\mathcal{A}:~~~~~ \sum_{i\in \mathcal{P}_j} x^c_{i|j}\leq 1
\end{equation}
\end{constraint}

\begin{constraint}[PAL Assignment per County]
The number of channels assigned to any PAL user in any county equals the number of channel licenses held by that PAL user for that county.  
That is,
\begin{equation}
\label{eq:constraint2}
    \forall i\in\mathcal{P}_j,\forall j\in\mathcal{A}:~~~~~
    \sum_{c\in\mathcal{F}_\mathrm{P}} x^c_{i|j}=L_{i|j}^P   
\end{equation}   
where $L_{i|j}^P$ is the number of licenses that PAL user $i$ holds in county $j$.
\end{constraint}

\begin{constraint}[PAL Interference Protection]
The aggregate interference received by a PAL user device on any channel, which is contributed by other PAL users and GAA users, should not exceed a threshold.    
That is,
\begin{equation}
\label{eq:constraint3}
\begin{split}
    \forall c\in\mathcal{F}_\mathrm{P},
    \forall l\in\mathcal{A},
    \forall k\in\mathcal{P}_l,
    \forall a\in\mathcal{B}_{k|l}:~~~~~~~~~~~~~~~~~~\\
    x_{k|l}\sum_{j\in\mathcal{A},j\neq l} \sum_{i\in\mathcal{P}_j,i\neq k} \sum_{b\in\mathcal{B}_{i|j}} x_{i|j}^c\cdot \gamma_{b,a}^c~+ ~~~~~~~ ~\\
    x_{k|l}\sum_{j\in\mathcal{A}} \sum_{i\in\mathcal{G}_j} y_{i|j}^c \cdot \delta_{i,a}^c \leq T_k^\mathrm{P}
\end{split}
\end{equation}
where $T_k^\mathrm{P}$ is the interference threshold for PAL user $k$, which we assume, without impacting our later design, to be uniform across all channels and counties. $\mathcal{B}_{k|l}$ is the set of devices of PAL user $k$ in county $l$; $\gamma_{b,a}^c$ is the received signal strength (RSS) from PAL user device $b$ on channel $c$, evaluated at a location in the Priority Protection Area (PPA) of device $a$ that is closest to $b$; $\delta_{i,a}$ is the RSS from GAA user $i$ on channel $c$, evaluated at a location in the PPA of PAL user device $a$ that is closest to $i$.  
\end{constraint}

\begin{constraint}[GAA Assignment]
The number of channels assigned to each GAA user in any county should not exceed a target number.  
That is,
\begin{equation}
\label{eq:constraint4}
    \forall i\in\mathcal{G}_j,\forall j\in\mathcal{A}:~~~~~
    \sum_{c\in\mathcal{F}}y_{i|j}^c\leq L_{i|j}^\mathrm{G}
\end{equation}
where $L_{i|j}^\mathrm{G}$ is such target number for GAA user $i$ in county $j$.
\end{constraint}

\begin{constraint}[GAA Mutual Interference Control]
Each pair of GAA users assigned to any common channel should be geographically separated so that they do not interfere with each other's transmission. That is,
\begin{equation}
\label{eq:constraint5}
\begin{split}
    \forall c\in\mathcal{F},\forall j\in\mathcal{A}, \forall a,b\in\mathcal{G}_j,a\neq b:~~~~~~~~~~~~~~~~~~~~~\\
    D_{a,b}\geq y_{a|j}^c \cdot y_{b|j}^c \cdot (R_a+R_b)
\end{split}
\end{equation}
where $D_{a,b}$ is the distance between GAA users $a$ and $b$; $R_a$ and $R_b$ are their transmission ranges, respectively. In SAS operation, $D_{a,b}$, $R_a$, and $R_b$ can be conveniently obtained from users' spectrum requests.
\end{constraint}

\begin{constraint}[Incumbent Interference Protection]
If there is an active incumbent, the aggregated interference from all PAL and GAA users on any incumbent-active channel in any DPA should not exceed a threshold. That is,
\begin{equation}
\label{eq:constraint6}
\begin{split}
    \forall c\in\mathcal{F},
    \forall m\in\mathcal{I}: ~~~~~~~~~~~~~~~~~~~~~~~~~~~~~~~~~~~~~~~~~~\\
    Z_m^c\sum_{j\in\mathcal{A}}\sum_{i\in\mathcal{P}_j}\sum_{b\in\mathcal{B}_{i|j}}x_{i|j}^c\cdot\alpha_{b,m}^c ~ + ~~~~~~~~~~~~ \\
    Z_m^c\sum_{j\in\mathcal{A}}\sum_{i\in\mathcal{G}_j}y_{i|j}^c\cdot\beta_{i,m}^c \leq T_m^\mathrm{I}~~~~~~
\end{split}
\end{equation}
where $T_m^\mathrm{I}$ is the interference threshold for all locations in DPA $m$. $Z_m^c$ is a binary indicator---it equals $1$ if channel $c$ is used by an active incumbent in DPA $m$. $\alpha_{b,m}^c$ is the RSS from PAL user device $b$ on channel $c$, evaluated at a DPA point on DPA $m$'s boundary that is closest to $b$; $\beta_{i,m}^c$ is the RSS from GAA user $i$ on channel $c$, evaluated in a similar manner.
\end{constraint}


\subsection{Zero-knowledge Proof}

Zero-Knowledge Proof (ZKP) is a general protocol between two parties that allows the prover to convince the verifier that it knows a secret (or ``witness'') for which a given statement is true without revealing any information beyond the truth of the statement. Zero-Knowledge Succinct Non-Interactive Arguments of Knowledge (zk-SNARK) \cite{bitansky2012extractable} is a type of ZKP system specifically designed for efficiently proving the validity of a computation. Theoretically, they rely on the Knowledge-of-Exponent assumption \cite{naor2003cryptographic}, which guarantees that any prover capable of producing a valid proof must computationally ``know'' the witness (not just that the witness exists). zk-SNARKs are characterized by their small proof size (\textit{succinctness}), asynchronous interaction (\textit{non-interactivity}), and sublinear verification time, regardless of the complexity of the underlying computation. 

Technically, a zk-SNARK is constructed as a modular combination of two layers: an information-theoretic layer, which compiles a computation into an arithmetic circuit and attests to its satisfiability, and a cryptographic layer, which enforces integrity via a polynomial commitment scheme and achieves non-interactivity using the Fiat-Shamir transformation \cite{chiesa2020marlin}. Together, a zk-SNARK allows a prover to prove knowledge of a witness $w$ for a public statement $x$ such that it satisfies the relation $C(x,w)=0$ for some arithmetic circuit $C$, without revealing $w$ (\textit{zero-knowledge}). The system guarantees that an honest prover can always convince the verifier (\textit{completeness}) and that a dishonest prover cannot convince an honest verifier that a false statement is true (\textit{soundness}).

In our design, we use the Groth16 \cite{groth2016size} implementation of zk-SNARK, mainly because of its verification time efficiency and availability of the supporting open-source JavaScript library \texttt{snarkjs} \cite{snarkjs}. Groth16 achieves the theoretical lower bound for proof size (3 group elements) and verification time (3 pairings + $n$ multiplications). This efficiency is achieved with the help of a circuit-specific trusted setup phase to generate the Common Reference String (CRS), which pre-computes a proving key linear in the circuit size and a constant-size verification key.
We use Circom \cite{circomdocs}, a specialized domain-specific language for creating arithmetic circuits, to convert the spectrum allocation policy constraints into Rank-1 Constraint Systems (R1CS). R1CS is a system of quadratic constraints that represents an arithmetic circuit over a finite field and handles equality and comparison constraints differently. We will describe how we build R1CS constraints from the validity conditions in our proofs.
\section{ZKPs for Allocation Constraint Validity}
\label{sec:sys-design-constraints}

In this section, we construct a ZKP system for checking the validity of allocation constraints described in Section \ref{subsec:list-constraints}. 


\subsection{ZKP Circuits for Allocation Constraints}

Our general strategy is to build a verification circuit that arithmetizes all constraints while maintaining data confidentiality. For all constraints, we treat the primary variables as \emph{private inputs} (i.e., the secret \emph{witness} of the proof), including the channel assignment indicators ($x^{c}_{i|j}$, $y^{c}_{i|j}$, and $Z_m^c$), per-user channel assignment limits ($L^{P}_{i|j}$ and $L_{i|j}^G$), RSS values ($\gamma_{b,a}^c$ and $\delta{i,a}^c$), distance values ($D_{a,b}$, $R_a$, and $R_b$), and interference thresholds ($T_k^\mathrm{P}$ and $T_k^\mathrm{P}$). This mitigates information leakage about the allocation to the verifier. 

\subsubsection{Constraints 1, 2, 4}
Constructions for these constraints are straightforward as they only involve integer operations.

For Constraint 1, as shown in Eq. (\ref{eq:constraint1}), the circuit enforces \emph{booleanity} of each $x^{c}_{i|j}$ and an \emph{at-most-one} condition per county, holding true if the summation equals $0$ or $1$.
For Constraint 2, as shown in Eq. (\ref{eq:constraint2}), 
the circuit enforces the \emph{cardinality constraint} that the summation equals to $L_{i|j}^P$
together with explicit \emph{range constraints} reflecting the system policy:
$\forall i:1 \le L^{P}_{i|j} \le 4 \quad \text{and} \quad \sum_{i\in \mathcal{P}_j} L^{P}_{i|j} \le 7$. These checks ensure that each PAL holder selects exactly as many channels as their license count permits, and thus satisfy the county-wide licensing limit specified by the policy.
For Constraint 4, as shown in Eq. (\ref{eq:constraint4}), 
the circuit enforces booleanity of each $y^{c}_{n\mid j}$ and cardinality constraint that the summation equals $L^{G}_{i|j}$,
together with the explicit range constraint reflecting the system policy:
$\forall i: 0 \le L^{G}_{i|j} \le 4$.
This guarantees that no GAA user exceeds their allowed target number of channel assignments.

\subsubsection{Constraints 3, 5, 6}
Constructions for these constraints are more complex as they involve handling fixed-point arithmetic and nested summations.

For Constraint 3, as shown in Eq. (\ref{eq:constraint3}), 
the circuit computes the total interference as a sum of gated contributions using coefficients $\gamma$ and $\delta$, and multiplies by the local activity indicator $x^{c}_{k|l}$ so that the bound is enforced only when the target PAL device is active on channel $c$. Since the computations are performed over a finite field, we represent real-valued coefficients and thresholds using \emph{decimal fixed-point} encoding: for a value $x \in \mathbb{R}$ we store the integer $\tilde{x} = \lfloor x \cdot 10^{f} \rceil$, where $f$ is the chosen number of decimal digits of precision.
Because circuit arithmetic is performed modulo a prime field, we apply range checks (bit-length bounds) to encoded inputs and to intermediate aggregates so that integer comparisons are well-defined and cannot be satisfied via modular wrap-around.

For Constraint 5, as shown in Eq. (\ref{eq:constraint5}), we observe that it enforces a minimum distance between any two GAA CBSDs assigned to the same channel $c$ within a county. 
For each unordered pair $(n,g)$, the circuit computes a conditional lower bound
$y^{c}_{n\mid j}\,y^{c}_{g\mid j}\,\bigl(R_{n\mid j}+R_{g\mid j}\bigr)$.
Distances and transmission ranges are represented as bounded integers using fixed-point encoding, and then an inequality between $D_{a,b}$ and the lower bound is implemented.

For Constraint 6, as shown in Eq. \ref{eq:constraint6},
For each DPA point $m$, the circuit aggregates the weighted PAL and GAA, and gates the inequality with the binary indicator $Z^{c}_{m}$, so the constraint is enforced only when $m$ is active on channel $c$.
Similar decimal fixed-point encoding and range limit checking techniques as with Constraint 3 are used.

\subsection{Additional Allocation Integrity Constraint}

With the channel assignment indicators used as private inputs to the proof, the proof has a weakened soundness property, since an dishonest prover can send fake channel assignments other than the ones used to generate the proof, to the clients. In other words, a verifier cannot verify whether the allocation provided to clients is the original one. To resolve this issue, we impose an additional constraint on our circuit construction to enforce the integrity of allocation, leveraging cryptographic commitments as public inputs.

\begin{constraint}[Allocation Integrity]
The channel assignments $\mathbf{alloc}$, as defined at the beginning of Section \ref{subsec:list-constraints} and treated as a private input (witness), should satisfy a public commitment according to the following constraint:
\begin{equation}
\label{eq:constraint7}
    \mathsf{com} = \mathsf{commit}(\mathbf{alloc})
\end{equation}
where $\mathsf{com}$ is a public input known to the verifier representing the precomputed commitment value; $\mathsf{commit}$ is the commitment function implemented within the arithmetic circuit.
\end{constraint}

The public commitment binds all allocation decisions to a single value, so that a prover cannot satisfy different constraints using inconsistent witnesses across separately generated proofs. 

To concretely implement Eq. (\ref{eq:constraint7}), we use a hash scheme $H:\mathbb{F}^{L}\rightarrow \mathbb{F}$ as the commitment function. $\mathbb{F}$ is a finite field and 
\begin{equation}
\label{eq:comm-input-size}
    L = A\cdot P + A\cdot G + M
\end{equation}
where $A=|\mathcal{A}|$ is the number of counties; $P$ and $G$ are the numbers of PAL users and GAA users, respectively (we assume they are fixed, without loss of generality); $M$ is the size of an auxiliary metadata vector $\mathrm{meta}$ that binds the commitment to the specific
problem instance (i.e., specifying $A$, $P$, $G$, and the channel index $c$), preventing replay of the same allocation commitment across different instances.

First, 
we define a flattened vector $\mathbf{v}$ as
\begin{equation}
    \mathbf{v}=\mathrm{Flatten}(\mathbf{alloc},\mathrm{meta}) \in \mathbb{F}^{L}
\end{equation}
where $\mathrm{Flatten}(\cdot)$ lists entries in the fixed order: $\{x\}$ in county-major and then user-major, followed by $\{y\}$ in county-major and then user-major, and followed by $\mathrm{meta}$. 
Finally, the commitment is computed by
\begin{equation}
    \mathsf{com} = H(\mathbf{v})
\end{equation}

\textbf{Realization with efficient hash.} We opt for Poseidon hash \cite{grassi2021poseidon} due to its arithmetic-native design, which drastically reduces the circuit size (number of R1CS constraints) compared to traditional hash algorithms like SHA-256. We provide the Poseidon hash configuration in Appendix \ref{app:poseidon}.

\subsection{Assembly of Proofs}




We generate a bundle of Groth16 proofs \cite{groth2016size} to certify that a single allocation witness $\mathbf{alloc}:=\{\{x\},\{y\}\}$ satisfies all constraint circuits, including the added Constraint 7. The proof generation consists of five steps:
\begin{itemize}
    \item 
\textbf{Step 1: Circuit compilation.}
For each constraint $k\in\{1,\dots,7\}$, we compile the corresponding Circom circuit into an R1CS representation and a WebAssembly (WASM) witness generator.

    \item
\textbf{Step 2: Trusted setup.}
We perform Groth16 setup in two phases: (i) a circuit-independent Powers-of-Tau ceremony, and (ii) a circuit-specific setup for each constraint circuit $k$ producing a proving key $.zkey$, a verification key $vk_k$, and the corresponding verification logic $\mathbf{Verify}_k$.

    \item
\textbf{Step 3: Witness generation.}
For each proof instance, we construct private inputs and compute the witness $w_k$ using the circuit's WASM witness generator. The public input includes $\mathsf{com}$.

    \item
\textbf{Step 4: Proof generation and bundling.}
Using the witness $w_k$ and proving key $.zkey$ for circuit $k$, the prover generates a Groth16 proof $\pi_k$ and public signal $public_k=\mathsf{com}$. The final allocation validity proof $\pi_{ACV}$ is written as:
\begin{equation}
    \pi_{ACV}:=\{(\pi_k,\text{public}_k,vk_k)\}_{k=1}^{7}
\end{equation}
\end{itemize}

\textbf{Verification.}
A verifier checks 
\begin{equation}
    \forall k: ~~~ \mathbf{Verify}_k(vk_k,\pi_k,public_k)=1
\end{equation}
and that all $public_k$ share the same commitment value $\mathsf{com}$. The verification is successful if all checks pass.

\textbf{Parallelization.}
To speed up proof generation, we use a two-stage execution strategy in our implementation: (i) a serial phase that performs per-constraint compilation, trusted setup, and witness generation, and (ii) a parallel phase that runs Groth16 proving across constraints (one process per constraint),
followed by off-chain parallel verification. Therefore, the wall time of the parallel proving phase is approximately determined by the slowest constraint prover thread, rather than the sum of all per-constraint proving times.

\subsection{Complexity Analysis}

Without the loss of generality, we assume each county has a fixed number $P$ PAL users (=$2$ in CBRS), a fixed number $N$ GAA users, $A=|\mathcal{A}|$ counties, the fixed number $F=|\mathcal{F}|$ channels (=$15$ in CBRS), and the length of the flattened commitment input vector is $L$, as defined in Eq. (\ref{eq:comm-input-size}). Table \ref{tab:analysis-complexity} shows the computational complexity associated with each constraint.
For Groth16, prover time is dominated by circuit size (number of R1CS constraints), which scales with the arithmetic and comparison/range-check operations in each circuit. As we will show in the experiments, 
Constraint 7 often dominates because its commitment has a large constant factor per permutation, so the end-to-end proving time is frequently close to the Constraint 7 proving time even though Constraint 7 scales only linearly in $L$.

In general, Groth16 verification is succinct: verifying one proof is independent of circuit size and costs a constant number of pairings plus a small linear term in the number of public inputs. Since our public inputs are small and per-proof verification
is treated as constant, verification scales with the number of proofs.

\begin{table}[ht]
    \centering
    \caption{Computational complexity associated with each constraint.}
    \begin{tabular}{cccc}
        \toprule
        Item  & Proof Gen & Proof File Size & Verification \\
        \midrule
        Constraint 1 & $\mathcal{O}(A)$             & $\mathcal{O}(1)$ & $\mathcal{O}(F)$ \\
        Constraint 2 & $\mathcal{O}(A)$             & $\mathcal{O}(1)$ & $\mathcal{O}(A)$ \\
        Constraint 3 & $\mathcal{O}(A^{2}\cdot N)$  & $\mathcal{O}(1)$ & $\mathcal{O}(F)$ \\
        Constraint 4 & $\mathcal{O}(A\cdot N)$      & $\mathcal{O}(1)$ & $\mathcal{O}(A)$ \\
        Constraint 5 & $\mathcal{O}(A\cdot N^{2})$  & $\mathcal{O}(1)$ & $\mathcal{O}(F)$ \\
        Constraint 6 & $\mathcal{O}(A\cdot N)$      & $\mathcal{O}(1)$ & $\mathcal{O}(F)$ \\
        Constraint 7 & $\mathcal{O}(L)$             & $\mathcal{O}(1)$ & $\mathcal{O}(F)$ \\
        \bottomrule
    \end{tabular}
    \label{tab:analysis-complexity}
\end{table}

\section{ZKPs for Move List Validity}
\label{sec:sys-design-movelist}





In practical SAS deployment, upon incumbent arrivals, computing a new allocation faces tight time constraints. Instead, the move list mechanism can be used to decide which grants should suspended so that incumbents receive below-threshold aggregate interference. 




\renewcommand{\algorithmicrequire}{\textbf{Input:}}
\renewcommand{\algorithmicensure}{\textbf{Output:}}
\begin{algorithm}
\caption{Standard move-list algorithm pseudocode \cite{souryal20183} augmented with \sysname's ZKP circuit generation steps (highlighted in \textcolor{blue}{blue} fonts)}
\label{alg:movelist}
\begin{algorithmic}[1]
\REQUIRE Protection channel $c$, protection threshold $t$, full set of protection points $\mathcal{PP}$, full set of grants $\mathcal{GR}$.
\ENSURE Move list for channel $c$ denoted $M_c$ ($\subseteq \mathcal{GR}$).

\STATE $M_c \leftarrow \emptyset$
\FORALL{point $p$ in set $\mathcal{PP}$}
    \STATE \textcolor{gray}{Find the set of grants in the ``neighborhood'' (within a certain radius) of protection point $p$ and channel $c$:} \\
    $G_{c,p} \leftarrow \mathrm{Neighborhood}(\mathcal{GR}, c, p)$;
    \STATE \textcolor{gray}{Sort grants in $G_{c,p}$ by their median interference contribution to point $p$ from smallest to largest:} \\ 
    $(g_1, \dots, g_n) \leftarrow \mathrm{SortAscending}(G_{c,p},\mathrm{MedInt}(G_{c,p}))$;
    \STATE \textcolor{blue}{Proof strategy: use the sorted list as a private witness and prove its sortedness (not the sort algorithm itself).}
    \FORALL{$a$ from $minAzimuth$ to $maxAzimuth$}
        \STATE \textcolor{gray}{Find the largest cutoff to the sorted list of grants such that the 95th percentile of the aggregate interference on $a$ does not exceed the protection threshold $t$:} \\
        $\hat{k}\leftarrow$ largest $k\in\{1,2,...,n\}$ s.t. \\
        \hspace{.5in} $\mathrm{AggInt95thprcntl}(\{g_1,...,g_k\},a)\leq t$; \\
        \STATE $\mathcal{M}_{c,p,a}\leftarrow\{g_{\hat{k}+1},...,g_n\}$;
        \STATE \textcolor{blue}{Proof strategy: generate a \emph{personalized proof per user}. The cutoff index $\hat{k}$ is kept as a \emph{private witness}, and we prove that it correctly satisfies the 95th-percentile threshold. Take the user’s grant identifier as a \emph{public witness} and prove that this identifier appears \emph{after} the cutoff in the sorted list, which justifies the public claim for that user $\texttt{is\_suspended}=1$.
}
    \ENDFOR
    \STATE $\mathcal{M}_{c,p}\leftarrow\bigcup_a \mathcal{M}_{c,p,a}$;
\ENDFOR
\STATE $M_c \leftarrow \bigcup_p M_{c,p}$;
\end{algorithmic}
\end{algorithm}

\subsection{ZKP Circuit Construction and Proof Assembly}

Algorithm \ref{alg:movelist} shows the pseudocode of the standard move-list algorithm according to \cite{souryal20183}, augmented with our proof components. The basic idea is to prove the validity of each important step of the algorithm, rather than rewriting the entire algorithm into circuits. Some complex components, like sorting and interference calculation, can be done ``off-circuit''; we directly use their outputs as ZKP witnesses.

Also, instead of proving the entire move list for all users (which is huge), we generate a proof personalized for each suspended user (i.e., those who are included in the move list).

The cutoff index $\hat{k}$, the threshold $T$, and the sorted interference list of all users of protection point $p$ and channel $c$ are treated as private witnesses. We first prove that the provided interference values are in nondecreasing order, so the user can trust that the list corresponds to a valid sorting outcome. Then the cumulative interference up to the cutoff is computed and the percentile condition: the aggregate value at $\hat{k}$ stays within the threshold, while adding the next element exceeds $T$, is proven, which establishes that $\hat{k}$ is the largest feasible cutoff. 
Finally, to enable personalized proofs, only the user’s grant identifier is stated as a public witness, and we prove that the user is suspended exactly when their position in the sorted list is greater than $\hat{k}$. This lets the verifier confirm suspension correctness without learning any information.

We follow the same proof assembly pipeline as with the previous section on allocation constraints. 
For each suspended user, we compile the circuit, run the Groth16 trusted setup, generate a witness from the private inputs, and then produce and verify the corresponding proof under the exported verification key.




\subsection{Complexity Analysis}


We assume that the size of the set of grants of a protection point $p$ and channel $c$ is $G$. Computation complexity for proof generation is $\mathcal{O}(G)$ since prover time is dominated by circuit size for Groth16. Also, as Groth16 verification does not scale with circuit size, computation complexity for proof verification is $\mathcal{O}(1)$.
\section{\sysname Use Cases}
\label{sec:usecase}





\sysname is designed to serve as a modular verifiability tool for general SAS paradigms. In this section, we describe two primary use cases compatible with existing models, followed by a new use case for next-era spectrum sharing.

\begin{figure}
  \centering
  \begin{subfigure}[b]{0.45\textwidth}
    \centering
    \includegraphics[width=\textwidth]{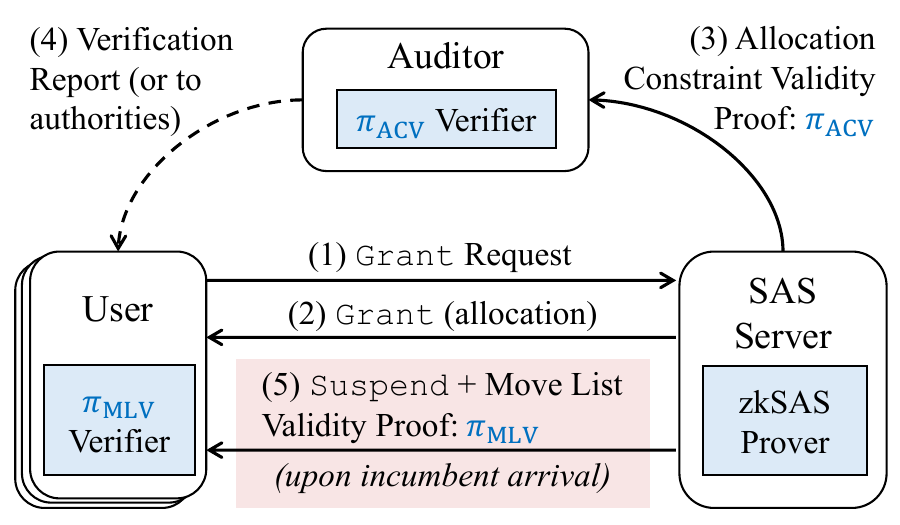}
    \caption{Use case 1: auditing centralized SAS functions.}
    \label{fig:usecase-existing-sas}
  \end{subfigure}

  \vspace{.1in}
  
  \begin{subfigure}[b]{0.45\textwidth}
    \centering
    \includegraphics[width=\textwidth]{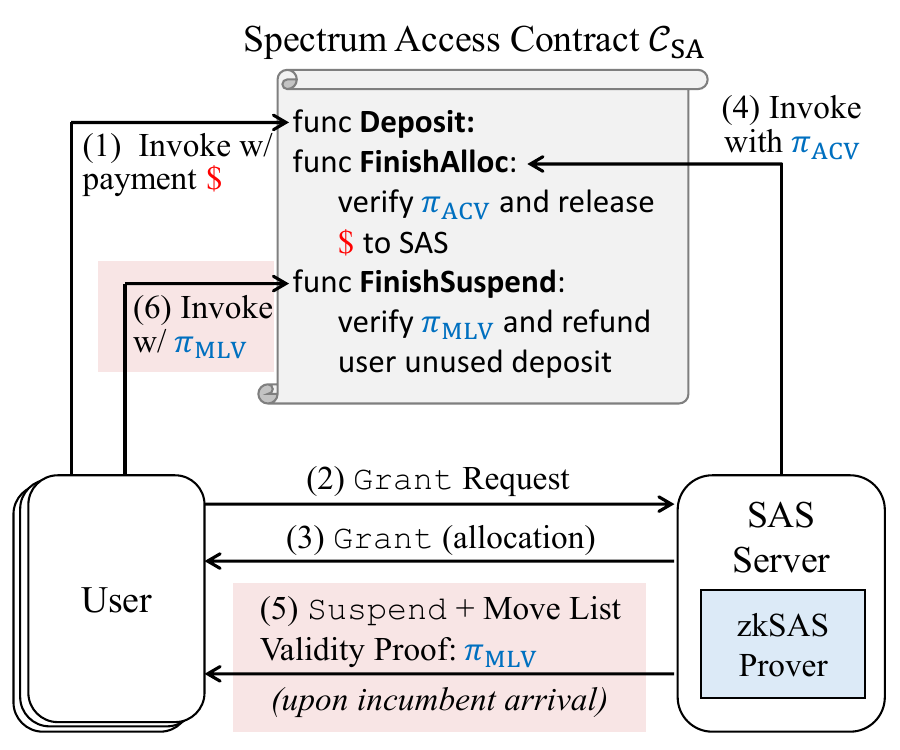}
    \caption{Use case 2: supporting blockchain-based SAS.}
    \label{fig:usecase-blockchain-sas}
  \end{subfigure}
  \caption{Two primary use cases of \sysname.}
  \label{fig:usecases}
  \vspace{-.1in}
\end{figure}

\subsection{Auditing Centralized SAS Functions} 

\sysname can be directly integrated into the current CBRS SAS implementation for checking the validity of grant generation and move-list inclusion. Fig. \ref{fig:usecase-existing-sas} shows an example workflow between a user, a SAS server, and an external auditor. After an initial inquiry (not shown in the figure), a user sends a \texttt{Grant} request to the SAS server (Step-1). The SAS server generates a new allocation, which comprises a list of grants for different users according to a certain algorithm, before sending the approved \texttt{Grant} back to each user (Step 2). The SAS server executes the \sysname prover to generate the allocation constraint validity proof $\pi_\mathrm{ACV}$ according to the method described in Section \ref{sec:sys-design-constraints} and sends it to the auditor for verification (Step 3), who routinely releases the verification report to users or a regulatory authority (Step 4). Notably, if there is no active incumbent during the allocation computation, $\pi_\mathrm{ACV}$ does not need to include proof of Constraint-6 for reduced overhead. 

When the SAS detects an incumbent arrival, it checks the up-to-date move list and sends the \texttt{Suspend} signal to users whose grants are in the move list (Step 5); this is accompanied by the move list validity proof $\pi_{MLV}$ produced according to the method described in Section \ref{sec:sys-design-movelist}. Upon verifying $\pi_{MLV}$, the affected user vacates the band immediately.

\subsection{Entrusting Off-chain Execution for Blockchain-based SAS} 


\begin{figure*}
  \centering
  \begin{subfigure}[b]{0.46\textwidth}
    \centering
    \includegraphics[width=\textwidth]{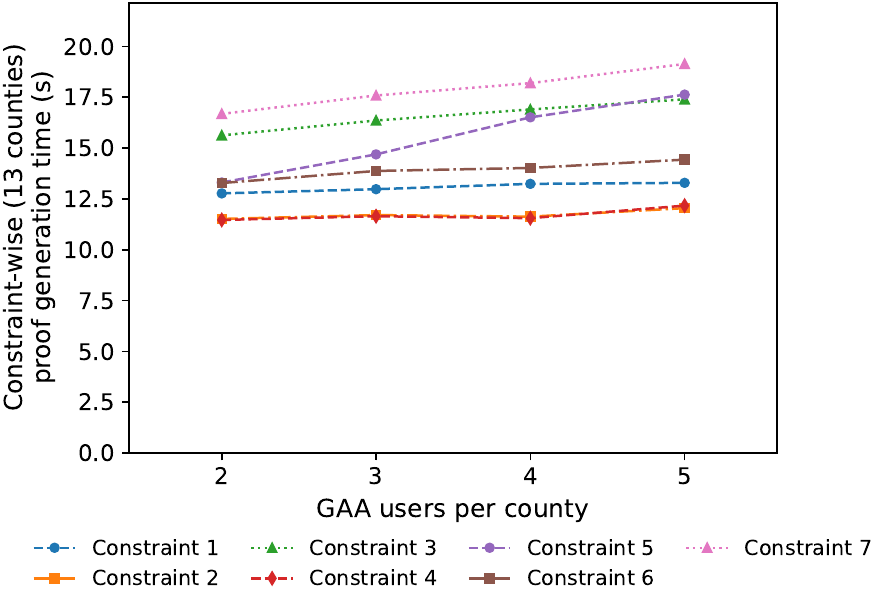}
    \caption{Constraint-wise proof generation}
    \label{fig:c-w-PG_13_P}
  \end{subfigure}
  \begin{subfigure}[b]{0.45\textwidth}
    \centering
    \includegraphics[width=\textwidth]{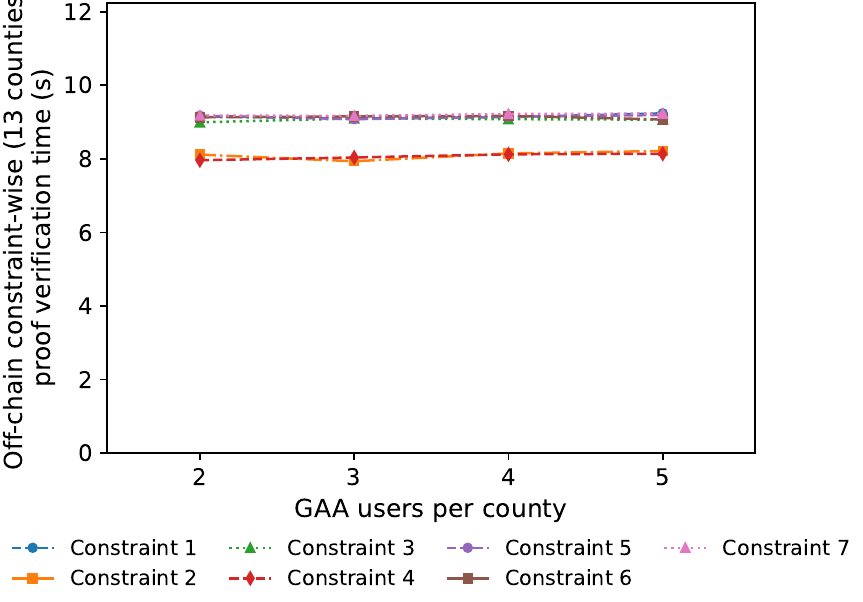}
    \caption{Constraint-wise proof verification}
    \label{fig:c-w-PV_13_P}
  \end{subfigure}
  \caption{Constraint-wise timings of allocation constraint validity proofs (Total County = 13) vs GAA users per county.}
  \label{fig:constraint-wise-13}
\end{figure*}

\sysname can be integrated into a general blockchain-based SAS framework to address the privacy and efficiency gap, as mentioned in Section \ref{sec:intro}, when coupled with \emph{off-chain execution}. Off-chain execution is an emerging computing paradigm in the blockchain/Web3 ecosystem that migrates an originally on-chain logic to an ``off-chain'' environment for private and efficient execution. Here, we use BD-SAS \cite{xiao2023bd}, a recent blockchain-based SAS framework, for example. The simplified workflow is shown in Fig. \ref{fig:usecase-blockchain-sas}. Instead of executing the grant request-allocation-payment logic entirely on the spectrum access contract $\mathcal{C}_{SA}$, we can offload the most computationally intensive step, allocation computation, to the off-chain server (an existing SAS server would suffice). A user initiates spectrum access by depositing the payment to $\mathcal{C}_{SA}$ (Step 1) and sending a \texttt{Grant} request to the off-chain server (Step 2). To claim the payment, the server needs to perform the allocation algorithm faithfully (Step 3) and upload the corresponding allocation constraint validity proof $\pi_\mathrm{ACV}$ to the contract, invoking the $\mathbf{FinishAlloc}$ function (Step 4). In an atomic step with this function, if $\pi_\mathrm{ACV}$ is verified, the payment will be released to the server. Invalid $\pi_\mathrm{ACV}$ will allow the user to claim back the deposit after a mandatory delay, which serves to mitigate request spamming from users (not shown in the figure).

Similar to the first use case, \sysname's move list validity proof $\pi_\mathrm{MLV}$ can facilitate grant suspension upon detecting incumbent arrivals (Step 5). Vacated users are entitled to claim back unused deposits (due to shortened grant period) by uploading $\pi_\mathrm{MLV}$ to the contract invoking the $\mathbf{FinishSuspend}$ function (Step 6). If $\pi_\mathrm{MLV}$ is verified, the unused deposit will be sent back to the user.


\subsection{Supporting ``Spectrum Bux'' in Next-gen Spectrum Sharing} 

To incentivize and monetize more efficient use of shared spectrum, next-generation spectrum sharing frameworks envision a market-based ecosystem where both incumbent and commercial users request short-term spectrum access by making Spectrum Bux (SBX) payments to a band manager \cite{nsf-newspectrum-solicitation}. In this model, SBX functions as a government-issued spectrum currency allocated to federal incumbent users, while Band manager, private-sector entitles analogous to SASs, provide allocation services. Aligned with this vision, band managers that provide Pay-As-You-Go (PAYG) services can leverages \sysname to generate validity proofs for real-time channel assignments, similar to use case 2. Furthermore, for service models requiring anonymous channel acquisition, such as those involving military users, the \sysname prover can be set up on the user side to generate zero-knowledge balance proofs. These proofs demonstrate sufficient SBX ownership, enabling band managers to authorize immediate service while deferring financial settlement. We will pursue these extensions in future work.


\section{Evaluation}
\label{sec:eval}


\begin{figure*}
  \centering
  \begin{subfigure}[b]{0.45\textwidth}
    \centering
    \includegraphics[width=\textwidth]{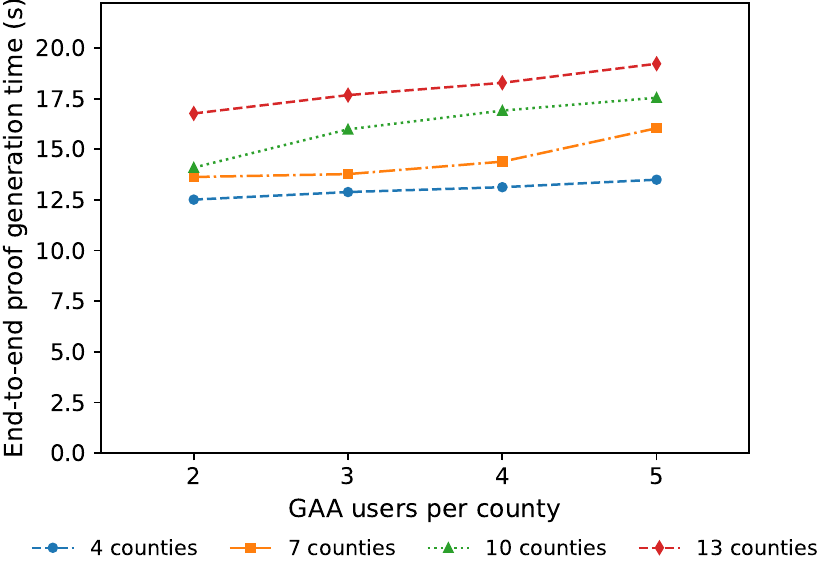}
    \caption{End-to-End proof generation}
    \label{fig:e-2-e-PG_13_P}
  \end{subfigure}
  \begin{subfigure}[b]{0.46\textwidth}
    \centering
    \includegraphics[width=\textwidth]{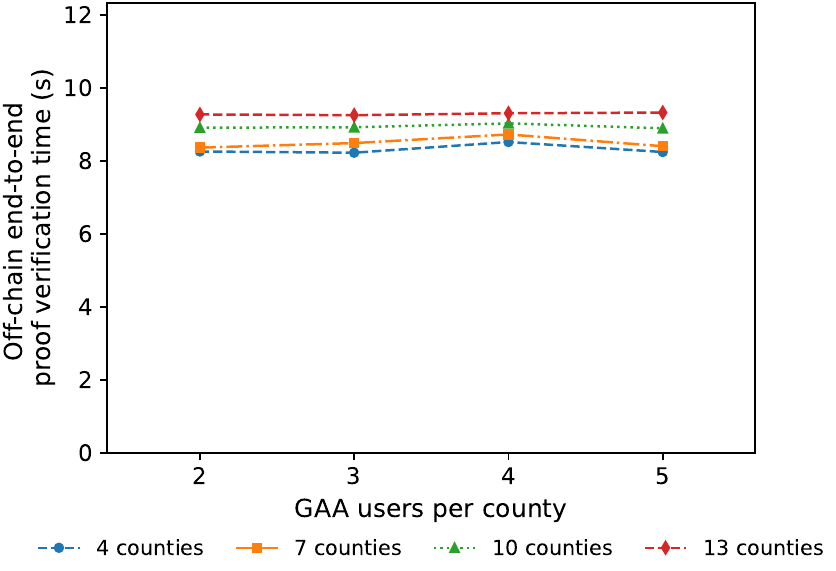}
    \caption{End-to-End proof verification}
    \label{fig:e-2-e-PV_13_P}
  \end{subfigure}
  \caption{End-to-End timings of allocation constraint validity proofs vs GAA users per county.}
  \label{fig:end-2-end}
\end{figure*}

\subsection{Implementation}


We implemented a prototype of \sysname using \texttt{circom} (v2.1.x) and \texttt{snarkjs} (Groth16 \cite{groth2016size}). Our implementation consists of the seven allocation-constraint circuits and the move-list validity circuit, totaling approximately 730 lines of circom code.\footnote{Source code and artifact available at:\\ https://github.com/purbasha-nishat/Spectrum-Allocation-Constraint-Circuits} For Constraint 5 specifically, we represented non-integer inputs using fixed-point encoding with one digit after the decimal point.

We tested our implementation on two platforms. Proof generations were tested on a MacBook Pro laptop with Apple M4 Pro and 24 GB of memory. Proof verification was tested on the same MacBook Pro laptop and a locally deployed blockchain system based on Hyperledger Fabric \cite{androulaki2018hyperledger}. Our Fabric environment ran on a single desktop server with Ubuntu 20.04 LTS and an Intel Xeon E5-2630 v3 @ 2.40\,GHz CPU (32 logical processors).

\subsection{Constraint-wise Computational Overhead}



We first evaluate the generation and verification overhead of our allocation constraint valid proof (Section \ref{sec:sys-design-constraints}) in our local laptop environment. The results are shown in Fig. \ref{fig:constraint-wise-13}. We observe that constraints~2 and~4 remain nearly constant as the number of GAA users per county increases. For the rest of the constraints, proof generation increases as the input size grows, but the growth rate varies widely by constraint. In particular, Constraint~5 exhibits the steepest increase, and it will become the dominant contributor to end-to-end proving time when the number of total counties and the number of GAA users per county increase.

Proof verification is comparatively stable across the same sweep, consistent with Groth16’s verifier cost being largely insensitive to witness size. However, verification for constraints 2 and 4 scales with the total number of counties, since the number of public witnesses grows with the total number of counties.




\subsection{End-to-end Computational Overhead}

Next, we evaluate the computational overhead of the holistic proof generation and verification process for both the allocation constraint validity proof and the move list validity proof.
Fig. \ref{fig:end-2-end} shows the results of allocation constraint validity proof. We observe that the end-to-end proof generation increases as the number of GAA users per county grows, and it also rises noticeably with the total number of counties (Fig. \ref{fig:e-2-e-PG_13_P}). This is expected because increasing either dimension expands the overall R1CS circuit size, causing the slowest constraint in the bundle to dominate the total proving time.
In comparison, end-to-end proof verification remains largely stable as the number of GAA users per county increases (Fig. \ref{fig:e-2-e-PV_13_P}), consistent with Groth16 verification being mostly insensitive to witness size. However, the verification curves shift upward as the number of counties grows, making verification primarily a function of county count rather than per-county GAA user count.

\begin{figure}
  \centering
  \includegraphics[width=0.4\textwidth]{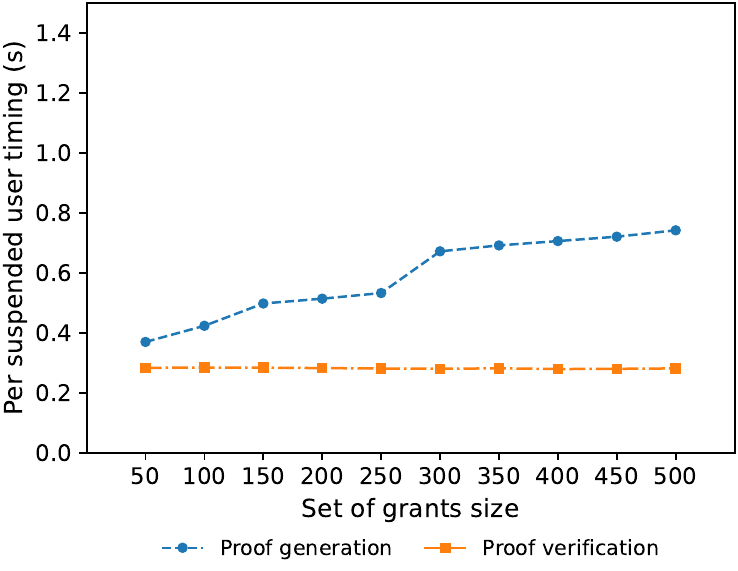}
  \caption{End-to-end timings of move list validity proofs per suspended user timings vs set of grants size.}
  \label{fig:movelist_PG_PV}
\end{figure}

The results of the move list validity proof are shown in Fig.~\ref{fig:movelist_PG_PV}. We observe that the proof generation time per suspended user increases as the grant-set size grows. In contrast, the verification time remains nearly constant across the same sweep, consistent with Groth16 verification being largely insensitive to the witness size.

\subsection{On-chain Verification Cost}

\begin{figure}
  \centering
  \includegraphics[width=0.4\textwidth]{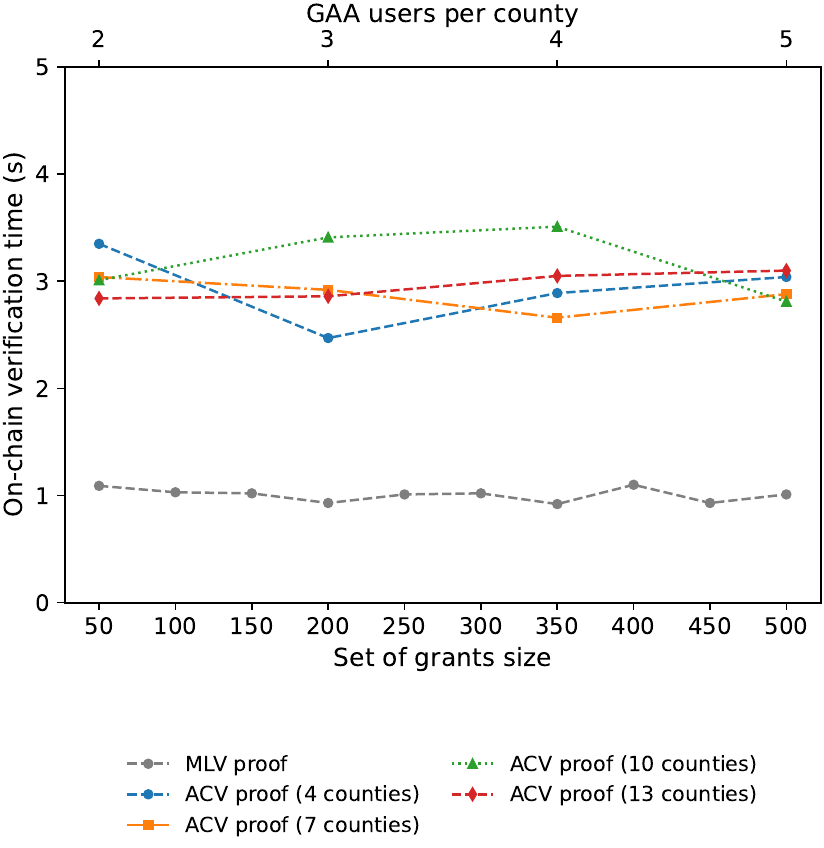}
  \caption{On-chain verification timings. ACV: allocation constraint validity, varying with GAA users per county (top 4 curves). MLV: move list validity, varying with set of grant size (bottom curve). }
  \label{fig:on-chain_merged_PV}
\end{figure}



We run the same verification steps on a Hyperledger Fabric (version 2.5.10) blockchain system with a dockerized network of 5 organizations, each with 1 peer node and 1 orderer node, deployed on a single server, a similar setup with BD-SAS \cite{xiao2023bd}. 

Fig. \ref{fig:on-chain_merged_PV} depicts the on-chain proof verification latency both for spectrum allocation constraints in different scenarios, and for the move list with a varying number of grants within a DPA. 
We measure the end-to-end latency for verifying the proofs of all constraints with varying numbers of counties and the number of users per county. The results show that the proof verification time is not affected by the SAS configuration or the complexity of the underlying constraints. The latency fluctuates between a range of 2.47 and 3.51 seconds, with no distinct upward trend correlated to the increase in the total number of GAA users. Notably, this on-chain verification takes shorter time than the off-chain counterpart (see Fig. \ref{fig:e-2-e-PV_13_P}) mainly due to two reasons---the blockchain's server platform is more powerful than the laptop and the blockchain network size is relatively small (5 validators and 5 orderers).

Similar to the allocation proofs, the move list verification demonstrates a constant-time performance profile. Despite the significant increase in the number of grants, from 50 to 500, the proof verification runtime cost remains stable around 1 second, which is one block cycle and the shortest possible time for finalization. This clearly demonstrates that an additional ZKP verification layer introduces a fixed overhead, regardless of the scale of interference in a DPA. In both cases, the results also demonstrate the constant verification time, a property of Groth16, regardless of the witness size.
\section{Related Work}

\subsection{Alternative ZKP Systems}
\label{subsec:alternative-zkp}

While our current implementation uses Groth16 \cite{groth2016size} to maximize on-chain verification efficiency, the underlying \sysname architecture is agnostic to the proof system and can be adapted to universal zk-SNARKs such as PlonK \cite{gabizon2019plonk} and Marlin \cite{chiesa2020marlin}. These protocols significantly reduce the burden of trusted setup by employing a single, updatable Structured Reference String (SRS) (instead of the per-circuit CRS in Groth16) usable across all circuits, with a marginal increase in prover overhead. Marlin \cite{chiesa2020marlin} also supports R1CS constraints natively, facilitating direct integration with our existing design.

Furthermore, \sysname may also be extended with transparent arguments that eliminate the trusted setup entirely, such as Bulletproofs \cite{bunz2018bulletproofs}, which relies on discrete logarithm assumptions; and STARKs \cite{ben2018scalable}, which utilize hash-based constructions to offer post-quantum security guarantees. However, adopting the transparent schemes for the current state of \sysname necessitates a different arithmetization or algebraic representation of the constraints and performance trade-offs: Bulletproofs incur linear verification time, while STARKs generally produce significantly larger proof sizes compared to pairing-based alternatives. As a future research direction, transparent zk-SNARK schemes can be leveraged to address spectrum management systems for diverse SAS operators.

\subsection{Blockchain-based Spectrum Access Management}


Ariyarathna et al. \cite{ariyarathna2019dynamic} utilize smart contracts to automate the dynamic leasing of spectrum resources while ensuring Service Level Agreements (SLA). It also incorporates a spectrum token design to tokenize frequency rights.
Zhang et al. \cite{zhang2020blockchain} propose a blockchain-based SAS that employs a ``Proof-of-Strategy'' mechanism, which integrates the computation of optimal spectrum allocation directly into the block validation process.
TrustSAS \cite{grissa2019trustsas} and BD-SAS \cite{xiao2023bd} are blockchain-based SAS designs that realize existing SAS functionalities, including channel inquiry, grant request handling, and grant issuance, with on-chain smart contracts. Both employ a two-level blockchain system design so that the global chain and the local chains handle different SAS tasks. TrustSAS further enables secondary users to query spectrum availability and commit to usage anonymously, while BD-SAS incorporates SAS server reshuffling to provide fault tolerance guarantees for each local chain.
TriSAS \cite{shi2024trisas} uses blockchain databases to realize a decentralized coordination process, as an alternative to the incumbent CPAS mechanism, to allow multiple independent SAS administrators to synchronize user inputs and agree on spectrum allocations without revealing their proprietary allocation algorithms. It further implements an on-chain verification module for checking the fairness and safety of proposed allocations.



\section{Conclusion}

We introduce \sysname, a practical zero-knowledge proof system for providing verifiability of spectrum management operations. \sysname can serve as standalone verification engine for existing SAS deployment to facilitate both external audits on allocation constraint satifiability and user audits on move list validity. It can be also integrated into blockchain-based SAS models coupled with off-chain execution to bridge the privacy and efficiency gap while establishing on-chain verifiability.
Our evaluation shows that \sysname achieves practical end-to-end latency at realistic county- and user-scale configurations, with verification remaining low-latency and proving time scaling predictably with instance size.
We envision \sysname as a valuable framework for establishing trust within next-generation spectrum sharing ecosystems. 


\section*{Acknowledgment}
This work was supported in part by the US National Science Foundation under grants 2433904, 2433905, 2331936, 232675, 2247560, 2247561; by the Office of Naval Research under grant N00014-24-1-2730; and by the Commonwealth Cyber Initiative.



\bibliographystyle{ieeetr}
\bibliography{reference,ref-standards}

\appendix
\subsection{Using Poseidon Hash}
\label{app:poseidon}

We use Poseidon hash \cite{grassi2021poseidon} in our allocation integrity proof (Constraint 7) with the following parameter configuration: $ARITY=16$ and $RATE=15$. $ARITY$ is the number of input elements the hash function can accept in a single chunk; $RATE$ is the number of field elements that are ``absorbed'' in each permutation of the hash function. Let
$$R=\left\lceil \frac{L}{RATE}\right\rceil$$
be the number of absorption rounds. For each round $r\in\{0,\dots,R-1\}$ and lane $j\in\{1,\dots,RATE\}$, we define, 
$$
b_{r,j}=
\begin{cases}
v_{r\cdot RATE + j} & \text{if } r\cdot RATE + j \le L,\\
0 & \text{otherwise.}
\end{cases}
$$
where $b_{r,j}$ denotes the $j$-th element of the $r$-th absorbed message block.  
Initialize the chaining state as $s_0=0$ and update it for each absorption round $r$ by
$$
s_{r+1} = \mathrm{Poseidon}\bigl(s_r, b_{r,1},\dots,b_{r,RATE}\bigr)
\quad \forall r\in\{0,\dots,R-1\}.
$$
After $R$ rounds, the final state $s_R$ defines the Poseidon-based hash output,
$$H(\mathbf{v}) := s_R$$
and the circuit enforces commitment consistency by requiring
$$\mathsf{com} = s_R$$

\end{document}